\documentclass[epj,referee]{svjour}
\usepackage{latexsym}
\usepackage{graphics}
\newcommand{\vk}{{\vec k}}
\newcommand{\vkp}{{\vec k}\,'}
\newcommand{\vq}{{\vec q}}
\begin{document}
\title{Anomalous dephasing of bosonic excitons interacting
with phonons in the vicinity of the Bose-Einstein condensation}
\author{O.M. Schmitt \inst{1}, D.B. Tran Thoai \inst{2}, 
P. Gartner \inst{3} and H. Haug \inst{1}}                     
\institute{Institut f{\"u}r Theoretische Physik, J.W.Goethe-Universit{\"a}t Frankfurt,
Germany 
\and permanent address: Institut of Physics, Ho Chi Minh City, Vietnam
\and permanent address: National Institut for Physics of Materials, Bucharest,
Romania }
\date{Received: date / Revised version: date}
%
\abstract{
The dephasing and relaxation kinetics  of bosonic excitons 
interacting with a thermal bath
of acoustic phonons is studied after coherent pulse excitation. 
The kinetics
of the induced excitonic polarization
is calculated within Markovian equations both for subcritical 
and supercritical excitation 
with respect to a Bose-Einstein condensation (BEC).
For excited densities $n$ below the critical density $n_c$,
an exponential polarization decay is obtained,
which is characterized by a dephasing rate $\Gamma=\frac{1}{T_2}$. 
This dephasing rate due to phonon scattering 
shows a pronounced exciton-density dependence in
the vicinity of the phase transition. It is well described
by the power law $\Gamma \propto (n-n_c)^2$ that can be 
understood by linearization of the equations around the
equilibrium solution.
Above the critical density we get a non-exponential relaxation 
to the final condensate value $p^0$ with 
$|p(t)|-|p^0|\propto\frac{1}{t}$ that holds for all densities.\\ 
Furthermore we include the full 
self-consistent Hartree-Fock-Bogoliubov (HFB) 
terms due to the exciton-exciton interaction and 
the kinetics of the anomalous functions 
$F_k=\langle a_k a_{-k}\rangle$. The collision terms are analyzed 
and an approximation is used
which is consistent with the existence of BEC.
The inclusion of the coherent x-x interaction does not
change the dephasing laws.  
The anomalous function $F_k$ exhibits a clear threshold 
behaviour at the critical density.
} 

\authorrunning{O.M. Schmitt et al.}
\titlerunning{Dephasing of excitons in the vicinity of $n_c$}
\maketitle

\section{Introduction}
Bose-Einstein condensation (BEC) is a fascinating topic 
attracting new interest, due to recent experimental 
observation for BEC of atomic ensembles \cite{Anderson,Davis} and 
evidence for BEC of excitons
in semiconductors  \cite{Mys,Butov,Lin,Goto}. 
In spite of the many similarities between
bosonic atoms and excitons in semiconductors
with large binding energy, the BEC of the latter has the interesting
specific feature  that the order parameter of the excitonic condensate is
identical to its optical polarization. 
Therefore it is directly accessible by
relatively simple optical measurements. 
This fact stimulated theoretical
predictions of unusual nonlinear optical properties in the
condensed regime \cite{Sham}.\\
In this paper the decay of the polarization $p$ (=order parameter)
is studied after excitation with a coherent laser beam, 
if one approaches the critical density $n_c$ for a BEC,
where an unusual dephasing kinetics can be expected. 
Unlike most studies of the condensation kinetics \cite{Laci,Ivanov,Tikhodeev},
here a large macroscopic occupation
of the condensate is created by the excitation process
via allowed optical transitions. This induced exciton amplitude
may then decay to its zero or nonzero stationary value.
The final stationary solution is perfectly described by 
the equilibrium theory of the free Bose gas.\\
One of the most promising candidates for an excitonic BEC
is $Cu_2O$ \cite{Snoke} with its extremely stable excitons. 
Due to its dipole forbidden exciton transition, an
experiment corresponding to our simulations would use  
two-photon transitions for the excitation of
ortho-excitons 
and a time-resolved polarization measurement 
via the same mechanism \cite{Frohlich}.
Although for a detailed quantitative simulation of such 
an experiment, additional
effects like polariton effects \cite{Claudia}, 
lifetime effects \cite{Tikhodeev} and the subtle excitation
mechanism will play a role even in the low density limit, 
our main predictions should be observable at least qualitatively
also in $Cu_2O$ close to a $BEC$.

\section{ Incoherent relaxation and dephasing kinetics for
excitons interacting with a bath of acoustic phonons \label{rate}}
1s-excitons with center-of-mass momentum $\vk$ are treated
in the boson approximation, i.e. the exciton operators 
fulfill the commutation relation  $\left[a_{\vk},a^{+}_{\vkp}\right]_{-}=\delta_{\vk,\vkp}$.
This approximation is only justified in semiconductors with large
exciton binding energy, when $na_0^{3}\ll1$ and mainly 1s-excitons are excited.
These excitons interact with longitudinal acoustic phonons
and a coherent classical light pulse via a dipole interaction.
The Hamiltonian is given by
\begin{eqnarray}
\label{hamil}
H&=&\sum _{\vk }e_{\vk }a_{\vk }^{\dagger}a_{\vk }+\sum _{\vq }\hbar \omega _{\vq }b_{\vq }^{\dagger}b_{\vq }\\ 
+&\frac{1}{\sqrt{V}}&\sum _{\vk ,\vq }g_{\vq }a_{\vk +\vq }^{\dagger}a_{\vk }(b_{\vq }+b_{-\vq }^{\dagger})
-\sqrt{V}\bigl( dE(t)a_{0}^{\dagger}+h.c.\bigr)\nonumber
\end{eqnarray}
with the dispersion of the excitons and phonons
\begin{equation}
e_{\vk }=\frac{\hbar ^{2}\vk ^{2}}{2m}\quad,\qquad \omega _{\vq }=c|\vq |\quad, 
\end{equation}
respectively, where $m$ is the translational exciton mass and
$c$ the sound velocity. 
The deformation potential matrix element
\begin{equation}
g_{\vq }=G\sqrt{\hbar \omega _{\vq }}
\end{equation}
in the long-wave length limit. The interaction constant $G$ is
given in terms of the deformation potential $D$ and the crystal 
density $\rho$ by
\( G^{2}=\frac{D^{2}}{\rho c^{2}} \). 
The finite photon momentum has been neglected.

We define further the exciton distribution function
$n_k=\langle a_k^{\dagger}a_k\rangle$ with $k\neq0$ and the exciton polarization
amplitude $p=\frac{1}{\sqrt{V}}\langle a_0\rangle$. 

One gets a closed set of kinetic equations for the 
polarization $p$ and the exciton occupation $n_k$,
by extending the standard method for deriving a semiclassical
Boltzmann equations also to the order parameter $p$ \cite{Laci}. 
The Heisenberg equations for $\langle a_0\rangle$ and $\langle a_k^{+}a_k\rangle$ are iterated to second order
in the interaction potential $g_q$. 
The higher order mean-values are factorized and the only correlations
kept are $p$ and $n_k$, e.g.
\[
\langle aa^{+}ab^{+}b\rangle\approx(1+\langle a^{+}a\rangle)\langle a\rangle\langle b^+b\rangle+\langle a\rangle\langle a^+a\rangle\langle b^+b\rangle.           
\]
Neglecting principal value contributions as
well as finite lifetime effects and
performing the Markov limit, one arrives at    
\cite{Laci}
\begin{eqnarray}
{\partial \over \partial t} n_{\vk}& =&
-\frac{1}{V} \sum_{\vkp}\left\{ W_{\vk \vkp}n_{\vk} 
(1+n_{\vkp}) - (\vk \rightleftharpoons \vkp )\right\}\nonumber\\
&-&\left[W_{\vk 0}n_{\vk} - W_{0\vk} (1+n_{\vk})\right ]|p|^2
\label{neq}
\\
{\partial \over \partial t} p& =&  
\frac{1}{2V}\sum_{\vkp} \left[W_{\vkp 0}n_{\vkp} - W_{0 \vkp} (1+n_{\vkp})\right ]p\nonumber\\
&+&\frac{i}{2\hbar}dE_0(t) \quad .
\label{peq}
\end{eqnarray}
The transition rates are given by Fermi's golden rule
\begin{eqnarray}
W_{\vk \vkp} &=&\frac{2\pi G^2}{\hbar}
|e_{\vk}-e_{\vkp}|\Big[{N}_{\vkp-\vk}
\delta(e_{\vk}-e_{\vkp}+\hbar \omega_{\vkp-\vk})\nonumber\\
&+&
\left(1+{N}_{\vk-\vkp}\right)  
\delta(e_{\vk}-e_{\vkp}-\hbar \omega_{\vk-\vkp})\Big]\quad ,
\label{trans_usual1}
\end{eqnarray}
where $N_k=\frac{1}{e^{\beta\hbar\omega_k}-1}$ is the thermal distribution of the phonons.
The delta-functions are broadened into Lorentz\\-ians according to \cite{Laci}.
In what follows (except in the last chapter)
we will study the consequence of these equations. 

\section{Laser pulse induced Bose Einstein condensation}
\begin{figure}[ht]
\resizebox{0.5\textwidth}{!}{  
  \includegraphics{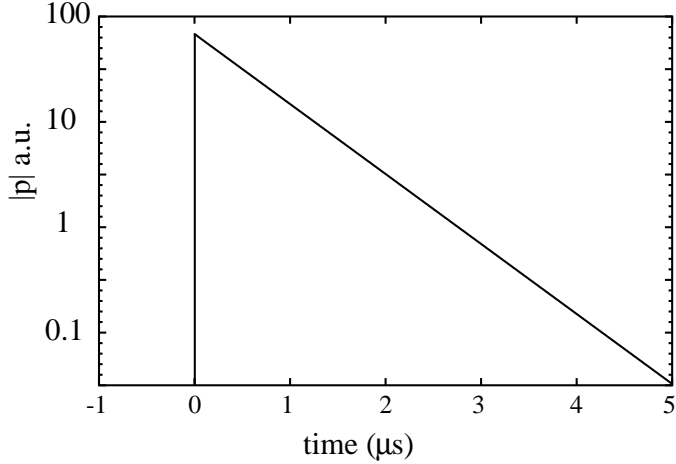}
}
\caption{Complete exponential dephasing for excitation with $n<n_c$}
\label{zerfall}
\end{figure}

\begin{figure}[ht]
\resizebox{0.5\textwidth}{!}{%
  \includegraphics{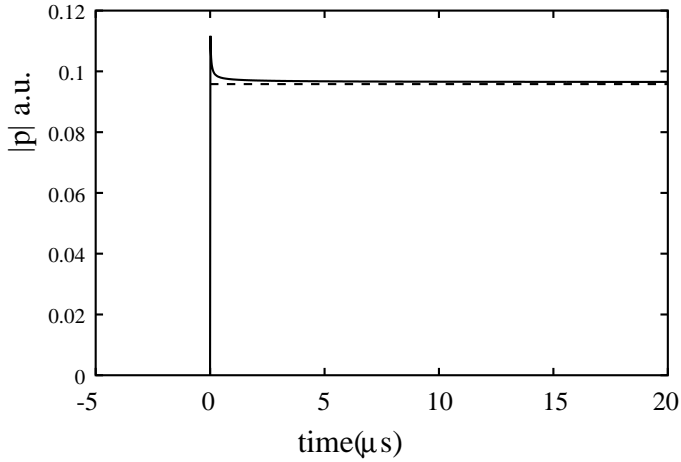}
}
\caption{Non-exponential approach of the exciton amplitude to condensation 
for excitations with $n>n_c$. Full line - calculated polarization,
dashed line - value predicted from equilibrium theory of the Bose-gas}
\label{cond}
\end{figure}

The strength of the laser pulse can be varied to excite densities
below (see Fig.\ref{zerfall}) and above (see Fig. \ref{cond}) 
the critical density for the BEC. The laser pulse is tuned
to the lowest 1s-exciton resonance and has a duration of $2.5ns$.
As an example $ZnSe$ parameters have been chosen at a low
temperature of $0.5K$. Although the BEC in $ZnSe$ may be unlikely,
this semiconductor with its dipole allowed excitons serves well 
to demonstrate our ideas,
which may hold for different experimental setups and
materials. 
  
In Fig. \ref{zerfall} and Fig. \ref{cond} a 
light pulse induced condensation is obtained. At excitation 
densities below
the critical density the polarization $|p|$ decays exponentially until
it vanishes as expected from usual optical experiments (Fig. \ref{zerfall}).
At stronger
excitations above the critical density, the exciton polarization 
approaches non-exponentially
its finite stationary value (Fig. \ref{cond}) , 
which is non-zero due to the onset of condensation.
This stationary polarization  is given
in terms of the excited density $n$ and the temperature $T$ 
in the form of $|p^0|=\sqrt{n-n_c(T)}$
as expected form the equilibrium theory of the Bose gas.
Obviously the finite lifetime of the excitons due to spontaneous
recombination will eventually cause an additional dephasing and
a decay of the exciton density.

\section{Critical slowing-down of the dephasing for approaching the critical
density from below}
\begin{figure}[ht]

\resizebox{0.5\textwidth}{!}{%
  \includegraphics{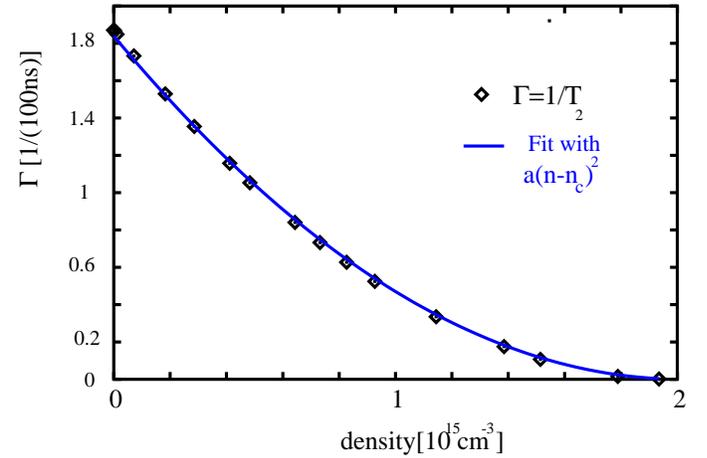}
}

\vspace{0.5cm}
\caption{Density dependence of the dephasing rate $\Gamma$ for $n_c=2\cdot10^{15} cm^{-3}$.}
\label{dephas}
\end{figure}

Below the critical density the decay of the polarization is
exponential, therefore a dephasing time $T_2$ and a dephasing rate
$\Gamma=\frac{1}{T_2}$ can be defined. We study the exciton-density 
dependence 
of these quantities as one approaches the critical density from
below. It is generally assumed  that the contribution of the
phonons to the dephasing rate is approximately  density-independent.
We can reproduce this result for moderate densities but do get
a considerable reduction of the dephasing rate near the critical 
density as shown in Fig.\ref{dephas}.
The density dependence of the dephasing rate can 
be fitted extremely well by a quadratic law
\begin{equation}
\Gamma \propto (n-n_c)^2,
\label{quad}
\end{equation}
which vanishes at the critical density, as
also shown in Fig. \ref{dephas}. 
This characteristic slowing-down of 
the exciton-dephasing kinetics can serve as a first 
experimental sign for the approach to the BEC.\\
A similar result has been observed also for exciton-exciton scattering
within similar rate equations \cite{oli} .

The quadratic law Eq.(\ref{quad}) can be derived
analytically by linearizing Eqs. \ref{neq} and \ref{peq} around
the stationary solutions 
$|p^0|$ and $n_k^0$. For the deviation $\delta p$ from the equilibrium value
$p^0=0$ one gets:
\begin{equation}
{\partial \over \partial t} \delta p =
\frac{1}{2V}\sum_{\vkp} \left[W_{\vkp 0}n_{\vkp}^{0} - W_{0 \vkp} (1+n_{\vkp}^{0})\right ]\delta p.
\end{equation}
Therefore one can immediately conclude that the final decay for $n< n_c$
is exponential and furthermore
\begin{equation}
\Gamma=\frac{1}{T_2}=\frac{1}{2V}(1-e^{\beta \mu})\sum_{\vk} W_{\vk 0} \frac{1}{e^{\beta e_k}-e^{\beta\mu}}. 
\label{power}
\end{equation}
On the other hand $(1-e^{\beta \mu})\propto(n-n_c)^2$ holds
as an exact thermodynamic relationship between 
the leading terms for
$\mu\approx0$ and $n\approx n_c$. However, the relation is relatively 
well obeyed even for large 
departures from the critical values, as can be checked by a simple
numerical calculation
Furthermore, due to the structure of the transition rates
Eq.(\ref{trans_usual1}), the main contribution in the sum of Eq.(\ref{power})
stems from the $k$-values away from the origin (the non-zero solution of
the delta-function). Therefore, for small temperatures the $\mu$-dependence
of the sum is weak and does not affect the $(n-n_c)^2$ behaviour of
its prefactor.

\section{Power law relaxation to the equilibrium condensate for $n>n_c$}
\begin{figure}[ht]
\resizebox{0.5\textwidth}{!}{%
  \includegraphics{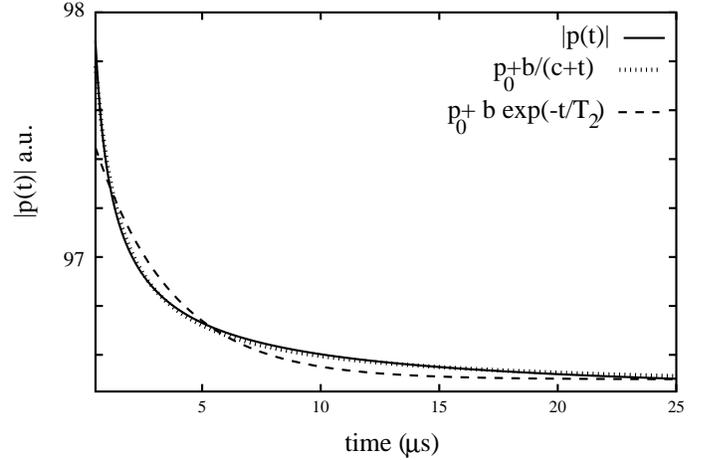}
}

\vspace{0.5cm}
\caption{Asymptotic kinetics of $|p(t)|$ for excitations with $n>n_c$,
power law fit $|\delta p(t)|\propto \frac{1}{t+c}$ dotted line. Exponential fit 
dashed line.}
\label{pot}
\end{figure}

While the polarization $|p|$ vanishes exponentially at subcritical
excitation, it reaches its stationary value $|p^0| =\sqrt{n-n_c} \neq 0$, 
when the excited density $n$ exceeds $n_c$. But in contrast to the subcritical
behaviour, this approach to equilibrium of the ideal Bose gas 
is not exponential any more. In contrast the dephasing kinetics 
is well described by a simple power law,
\begin{equation}
|p(t)| =|p^0|+a/(t+b), 
\end{equation}
as shown in Fig. \ref{pot}.

The slow non-exponential approach to equilibrium
for $n>n_c$ would also give in experiments sensitive to coherence 
a clear signature for the BEC, which may be
observable even when the final condensate is small and 
cannot be detected clearly.

Above the critical density the same linearization procedure like the one 
used in the last section but with $\mu=0$, $p=p^0+\delta p$ and $p^0\neq0$
yields
\begin{equation}
\frac{\partial}{\partial t} \delta p = \frac{1}{2V} \sum_{\vk} (1-e^{-\beta e_k})W_{k0} \delta n_k p^0 \quad .
\end{equation}
As shown in Ref. \cite{Laci}, the long-time behaviour for $\delta n_k$
is not exponential. The spectrum of the linear operator
controlling the $t\to\infty$ asymptotics of $\delta n_k$ is continuous and starts
from zero. Therefore, it is expected that a power-law decay is obeyed,
which is in accordance with our numerical results.

\section{Combination of phonon-scattering with HFB-correlations} 
The model used in the last sections may be useful
to describe the incoherent dephasing and relaxation phenomena
of the light-induced BEC. However, it does not account for a variety
of interesting and important phenomena, where exciton-exciton
correlations play an important role. These excitonic correlations are 
known to be important for the nonlinear
optical response of semiconductors \cite{Axt,Koch,Sham2}.
Additionally, essential aspects of the BEC can be only understood
in terms of an interacting Bose-gas.
These phenomena are e.g. superfluidity \cite{Mys,Butov}, 
the renormalized spectrum of the elementary excitations 
and some interference phenomena based
on the Gross-Pitajevski equation.
These effects are taken into account by the self-consistent 
Hartree-Fock-Bogoliubov treatment of the boson-boson interaction
that is also called  Girardeau-Arnowitt approximation. 
The impact of this approximation on the BEC 
has been reviewed e.g. in Refs. \cite{Hohenberg}.
We will now give a guideline how to include the HFB approximation in
the phonon scattering kinetics in such a way, that the results of
the last sections are conserved, but moreover coherent aspects
are taken into account. 

In the following the derivation of these equations is described.

To the Hamiltonian Eq. (\ref{hamil}) 
the exciton-exciton interaction $H'$ with a contact potential $W$ is added
\begin{equation}
H'= \frac{1}{4V}\sum_{\vec{k}_{1},\vec{k}_{2},\vec{q}}\!\!
W
a_{\vec{k}_{1}}^{+}a_{\vec{k}_{2}}^{+}
a_{\vec{k}_{1}+\vec{q}}a_{\vec{k}_{2}-\vec{q}}.
\end{equation}
The interaction matrix element $W$ between the excitons is in general 
given by a momentum-dependent expectation value of the Coulomb interactions
between the various point charges in the two excitons containing both the
attractive direct and the repulsive exchange interaction.
In a low-temperature exciton system the small momentum transfer dominates. In
this limit the direct interaction can be neglected 
and the exchange integral can be calculated analytically
\cite{263}. This yields a repulsive contact potential with
the coupling constant
\begin{equation}
W=\frac{26}{3}\pi a_0^3E_r.
\end{equation}  
Here $a_0$ and $E_r$ are the Bohr radius and the Rydberg energy, 
respectively. Note however, that the possibility of a biexciton formation is
lost in this long-wavelength approximation.  

In the HFB approximation all possible contractions have to
be taken which conserve the one-particle structure of the
Hamiltonian and its translational invariance. 
Therefore we have to introduce additionally to $p$ and $n_k$
the anomalous function $F_k(t)=\langle a_k(t)a_{-k}(t)\rangle$.
From the viewpoint of the underlying e-h picture, the bosonic HFB
approximation includes also the dynamics of four-particle correlations
(two electron and two hole operators). Therefore, correlations
beyond the usual electronic HF approximation are taken into account.

\bigskip
To derive the kinetics, the 
Heisenberg equations are iterated to first order in $W$ 
and to second order in $g$. 
The mean-values of higher orders are then factorized as
explained in section \ref{rate},
only that now all possible contractions are taken into account 
including the anomalous function $F_k$. 
The phonon part(collision terms) is 
treated in the Markov approximation and contributions 
from principal value integrals have been neglected. 
Due to the fact that the number of
possible second-order phonon contributions is quite large,
all anomalous functions $F_k$ are neglected in the collision terms. 
We will show later that
the contributions of the anomalous functions to the collision terms within
the same approximation scheme make the condensate unstable.
In the language of diagram theory, our approximation
includes the HFB diagrams \cite{oli} 
and a selection of second-order phonon scattering
diagrams.\\ 
The resulting equations are 
\begin{eqnarray}
\frac{\partial }{\partial t}n_{k}&=&\frac{W}{\hbar }\; \Im \: \left\{ \left( p^{2}+\frac{1}{V}\sum _{\vq}
F_{q}\right) F^{*}_{k}\right\} \nonumber\\
&+&\frac{\partial }{\partial t}n_{k}|_{coll} \label{ncoh}\\
\frac{\partial }{\partial t}F_{k}&=&-\frac{i}{\hbar }\left[ 2e_{k}+2W(n_{0}+\frac{1}{V}\sum _{\vq}n_{q})\right] F_{k}\nonumber\\
-\frac{i}{2\hbar }\big( 2n_{k}&+&1\big) W\left( p^{2}+\frac{1}{V}\sum _{\vq}F_{q}\right) +\frac{\partial }{\partial t}F_{k}|_{coll} \label{Fcoh}\\
\frac{\partial }{\partial t}p&=&-\frac{i}{\hbar }\left[ e_{0}+\frac{W}{2}(n_{0}+\frac{2}{V}\sum _{\vq}n_{q})\right] p \nonumber \\
-\frac{i}{2\hbar }&W&\, p^{*}\frac{1}{V}\sum _{\vq}F_{q}+\frac{i}{2\hbar }d\: E+\frac{\partial }{\partial t}p|_{coll}\label{pcoh} 
\end{eqnarray}

\begin{eqnarray}
\frac{\partial }{\partial t}n_{k}|_{coll}&=&-\frac{2\pi }{\hbar V}\sum _{\vq}g_{\vk-\vq}^{2}\{\delta(e_{q}-e_{k}-\hbar \omega _{\vk-\vq})
\nonumber\\
\times\big[ N_{k-q}n_{k}(1+n_{q})&-&(N_{k-q}+1)(1+n_{k})n_{q}\big] -(\vk \leftrightarrow \vq)\} \nonumber\\
&+&\frac{2\pi }{\hbar }g_{k}^{2}\delta (e_{k}-\hbar \omega _{k})n_{0}(N_{k}-n_{k})
\label{ncol}
\end{eqnarray}
\begin{eqnarray}
\frac{\partial }{\partial t}p|_{coll}&=&-\frac{\pi }{\hbar V}p\sum _{\vq}g_{q}^{2}\: \delta (e_{q}-\hbar \omega _{q})(N_{q}-n_{q})
\label{pcol}\\
\frac{\partial }{\partial t}F_{k}|_{coll}&=&-\frac{2\pi}{\hbar}g_{k}^{2}\delta(e_{k}-\hbar\omega_{k})\left[ p^{2}(N_{k}-n_{k})\right]
\label{Fcol}
\end{eqnarray}

\begin{figure}[ht]
\resizebox{0.5\textwidth}{!}{%
  \includegraphics{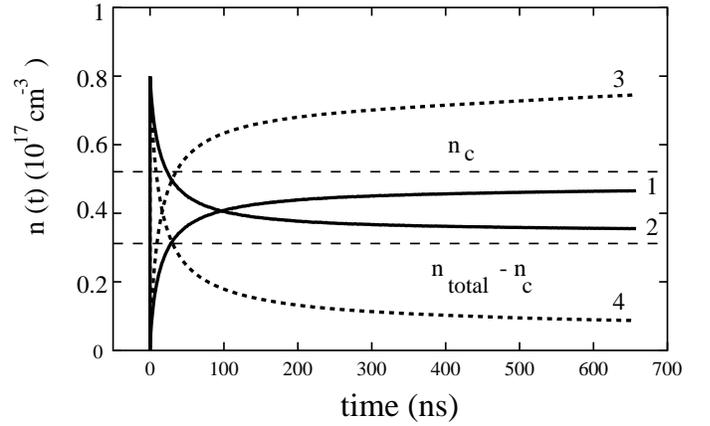}
}
\vspace{0.5cm}
\caption{Density $n$ and $n_0$ versus time $t$ above the critical
density with $n_{total} = 8.3 \cdot 10^{16} cm^{-3}$ and $T_{ph}=1.5K$.
Curve 1 and 2: $F$ neglected($F^0$) in collision terms. Curve 3 and 4: 
first order contributions($F^1$) included in the collision terms.
Horizontal lines are $n_c=5.2 \cdot 10^{16}$ and $n_{total}-n_c=3.1\cdot10^{16} cm^{-3}$
respectively.} 
\label{F}
\end{figure}

First we demonstrate that the addition of scattering
terms proportional to $F_k$ in the collision terms of $F_k$ Eq.(\ref{Fcol}) 
destroy the condensate, while
omitting them leads to a stable condensate solution. 
In Fig. \ref{F} a solution of the equations is shown
above the critical density. While the coherent density $n_0=|p|^2$ goes
to zero if the collision terms proportional to $F_k$ ($F^1$-terms) 
are included (see curves 3,4), 
it condenses if only terms without $F_k$($F^0$-terms) are
considered as done in Eq.(\ref{Fcol}) (see curves 1,2).
For subcritical excitation the condensate decays in both approximations.

\smallskip
The reason for the numerically obtained condensate 
destroying properties of the anomalous
contributions in the collision terms are not yet fully understood.
It can be easily seen analytically that these terms are not in accordance
with the condensate solutions of the free Bose-gas.
However they are obviously also not yet fully compatible
with the condensation-solution of the interacting Bose-gas. More refined
approximations like the inclusion of energy renormalizations in 
the collision integrals are needed. An argument for omitting the contributions
of the anomalous functions to the second order scattering terms
has been proposed in Ref. 
\cite{Proukakis}. Here it is argued that an adiabatic elimination
of $F_k$ yields for a scattering term containing one anomalous function
a term of third-order $\propto Wg_q^2$, which is inconsistent
with the considered second-order scattering. Although this argument is not
rigorous in a self-consistent theory, all collision
terms proportional to $F_k$ are disregarded in the following.

\begin{figure}[ht]
\resizebox{0.5\textwidth}{!}{%
  \includegraphics{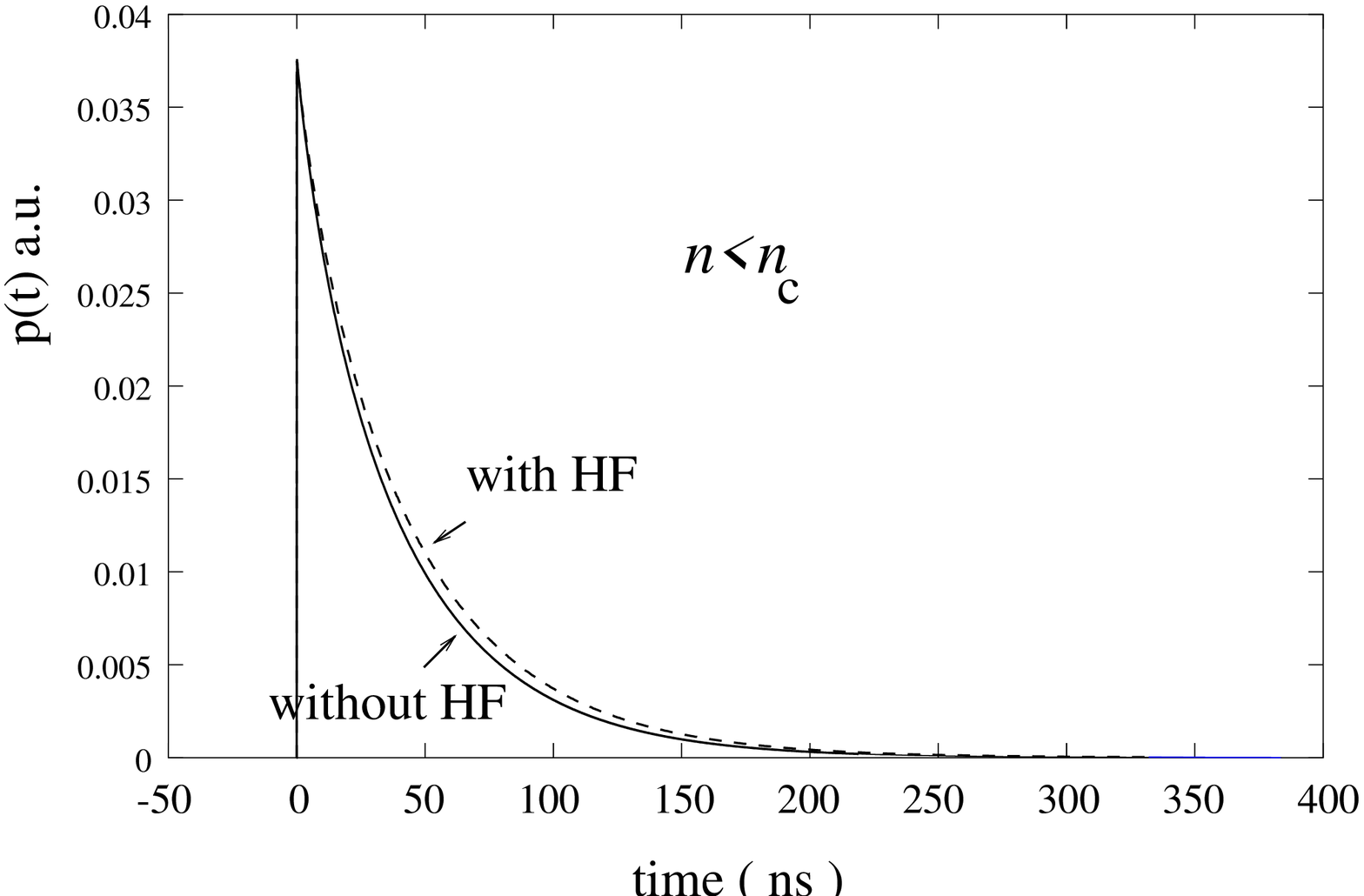}
}
\caption{Excitation at $n<n_c$, comparison HF, without HF}
\label{comp1}
\end{figure}

\begin{figure}[ht]
\resizebox{0.5\textwidth}{!}{%
  \includegraphics{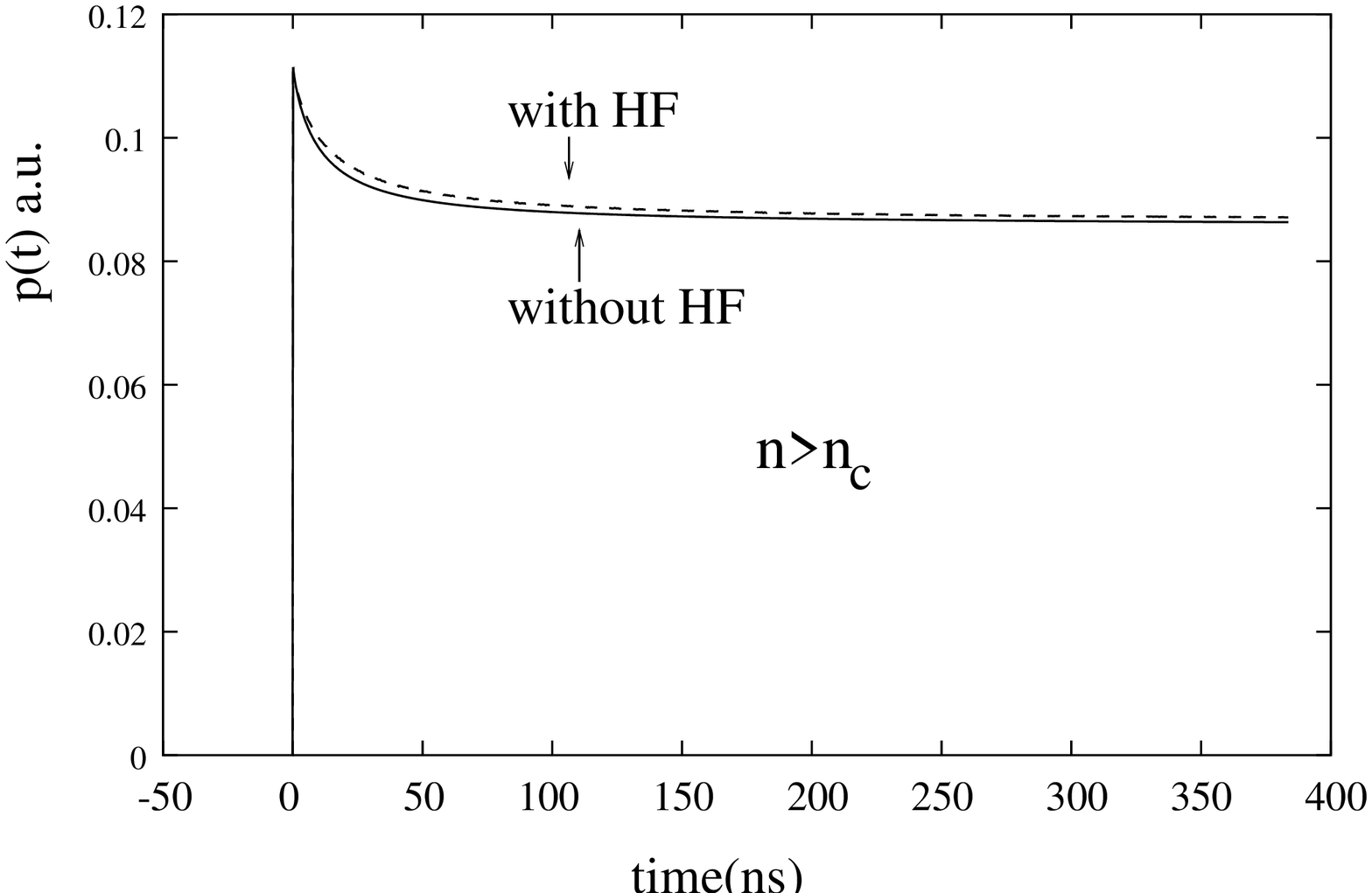}
}
\caption{Excitation at $n>n_c$, comparison HF, without HF}
\label{comp2}
\end{figure}

\bigskip
We compare the prediction for the kinetics of the pure phonon model 
Eqs.(\ref{neq}),(\ref{peq})
and the full model Eqs.(\ref{ncoh}-\ref{Fcol}) 
in two figures once below (Fig. \ref{comp1}) 
and once above (Fig. \ref{comp2})  the
critical density.

\begin{figure}[ht]
\resizebox{0.5\textwidth}{!}{%
  \includegraphics{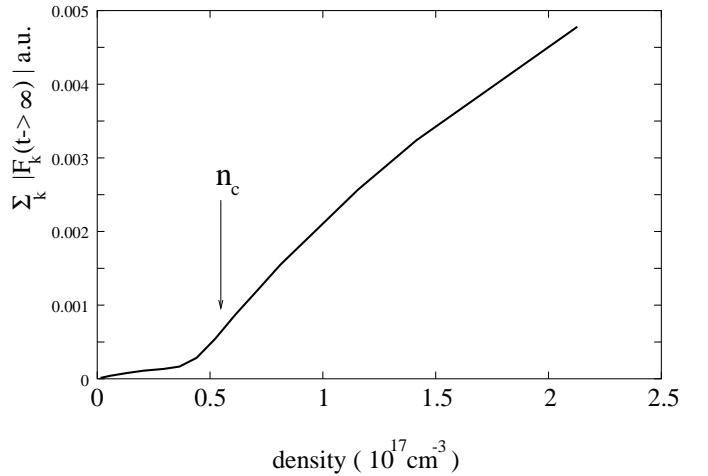}
}
\caption{Asymptotic sum of the anomalous function $F_k$}
\label{sum}
\end{figure}

The results of the two models concerning dephasing and relaxation
 can hardly be distinguished. 
We conclude that all our results in the pure
phonon model are stable 
against Hartree-Fock-Bogoliubov correlations.\\ 
Additionally the introduced anomalous function $F_k$ shows 
a threshold behaviour as expected in a phase transition.
This is shown in Fig. \ref{sum} where we plotted
$\sum_k|F_k(t=\infty)|$ versus the excitation density.

To conclude this chapter, we presented a
model which does show the well-understood dephasing and 
condensation properties of the earlier introduced simpler
boson model, but additionally can be applied to experiments
where the coherence plays a more important role than in simple
studies of the dephasing. 

\section{Conclusion}
It was demonstrated that the dephasing of an exciton
polarization due to phonon
scattering can have surprising features, when the BEC is approached.
For subcritical excitation with a laser pulse, the polarization dephasing
rate slows down with increasing density 
in the vicinity of the BEC with a quadratic power law.
For supercritical excitation the expected exponential dephasing changes to
a power law relaxation approaching a finite value. 
We hope that these results may stimulate corresponding 
experimental investigations.
In the last chapter it was shown 
that condensation kinetics is very sensitive to the choice
of specific scattering terms (diagrams). A model is constructed which includes
both the incoherent scattering due to phonons and the HFB correlations.
It has the same incoherent properties as the pure phonon model, but
may be also applied to experiments, where coherent correlations are important. 
\begin{acknowledgement}
This work has been supported by the DFG in the framework of the
Schwerpunktprogramm {\it Quantenkoh{\"a}renz in Halbleitern}.
\end{acknowledgement}

\end{document}